\documentclass[final,5p,times,twocolumn]{elsarticle} 

\usepackage{graphicx}
\graphicspath{ {pics/}}

\usepackage{threeparttable}

\usepackage[caption=false]{subfig}

\usepackage{lineno, hyperref}
\modulolinenumbers[5]

\usepackage{amssymb}
\usepackage{amsmath}
\usepackage{amsthm}
\usepackage{amsfonts}

\usepackage{braket}

\usepackage{hyperref}
\usepackage{array}
\begin{document}

\begin{frontmatter}

\author[a1]{Stefano Marin \corref{cor1}}
\ead{stmarin@umich.edu}

\author[a2]{Vladimir A. Protopopescu}

\author[a3,a4]{Ramona Vogt}

\author[a5]{Matthew J. Marcath}

\author[a1]{Stephan Okar}

\author[a1]{Michael Y. Hua}

\author[a5]{Patrick Talou}

\author[a1]{Patricia F. Schuster}

\author[a1]{Shaun D. Clarke}

\author[a1,a7]{Sara A. Pozzi}

\address[a1]{Department of Nuclear Engineering and Radiological Sciences, University of Michigan, Ann Arbor, MI 48109, USA}

\address[a2]{Oak Ridge National Laboratory, Oak Ridge, TN, USA}

\address[a3]{Physics Division, Lawrence Livermore National Laboratory, Livermore, CA}

\address[a4]{Physics Department, University of California, Davis, CA, USA}

\address[a5]{Computational Physics Division, Los Alamos National Laboratory, Los Alamos, NM 87545, USA}

\address[a7]{Department of Physics, University of Michigan, Ann Arbor, MI 48109, USA}

\title{Event-by-Event Neutron-Photon Multiplicity Correlations in $^{252}$Cf(sf)}
\date{\today}


\begin{abstract}
Excited nuclear fragments are emitted during nuclear fission. The de-excitation of these fission fragments takes place as sequential emission of neutrons followed by photons. A correlation between neutron and photon multiplicities accompanying fission is thus expected. Fission event generators based on established statistical nuclear physics models predict a negative event-by-event correlation in neutron-photon multiplicity. A survey of published experimental results of an event-by-event covariance between the neutron and photon multiplicities emitted following the spontaneous fission of $^{252}$Cf is presented.  Analytic unfolding expressions are developed in this work to determine the bias introduced by background sources, particle misclassification, pulse pileup, and inelastic photon production. The published experimental data are re-analyzed using these unfolding techniques and are found to be in qualitative agreement with the predictions of model-based calculations. In particular, we have concluded that there exists a significant event-by-event neutron-photon emission competition following the spontaneous fission of $^{252}$Cf.
\end{abstract}

\end{frontmatter}

\section{Introduction}
\label{sec:intro}

Since its discovery eighty years ago, nuclear scientists have been investigating the properties of the fission process. Due to the complexity introduced  by the large number of nucleons and strong forces involved, a complete and satisfactory description of fission is still lacking. Recently, fission event generators have been developed to model the complex behavior of fission fragments from scission to de-excitation. These generators apply established nuclear physics models to simulate the de-excitation of the fission fragments by evaporation of neutrons and photons. The predictions of the fission event generators have not been thoroughly experimentally validated. Spontaneous fission offers a simpler problem to analyze than induced fission because no additional particles are introduced. Due to its widespread use as a calibration source and its high spontaneous fission rate, $^{252}$Cf is a baseline isotope for fission studies. However, there are still many aspects of the spontaneous fission of $^{252}$Cf that are not fully characterized or understood, most notably the correlation between the emission of neutrons and photons following fission \cite{Talou2018}.

Our objective is to determine the neutron-photon emission correlation in the form of an event-by-event multiplicity covariance in the spontaneous fission of $^{252}$Cf. The neutron-photon multiplicity correlation has garnered significant interest over the past decades because of its connection to the energy-spin correlation in the formation of fission fragments \cite{Talou2018, Nifenecker1972, Grover1967b}. However, experimental data are scarce and inconclusive. Furthermore, the published event-by-event results \cite{Glassel1989, Bleuel2010, Marcath2018} have been found to be inconsistent with one another.

Several new and necessary results are presented in this work. We have analyzed and interpreted previous experimental determinations of event-by-event neutron-photon multiplicity correlations, and we have resolved the apparent inconsistencies between the various experimental determinations. Furthermore, we have resolved the disagreement between the experimental results and the model-based calculations. The corrections of the experimental results were made possible by the novel unfolding techniques we have developed. The analytic form of these techniques is presented here. 

The outline of this work is summarized as follows. Section~\ref{sec:back} introduces fission event generators and their predictions and reviews the published literature for experimental results. Section~\ref{sec:unf} presents the multiplicity-correlation unfolding techniques we have developed. The techniques are then used in Section~\ref{sec:exp} to remove experimental biases from published experimental data. The results of this analysis demonstrate that the inconsistencies between results in Refs.~\cite{Glassel1989, Bleuel2010, Marcath2018} are resolved. Lastly, we draw the main conclusions from this investigation and present possible avenues of future work in Section~\ref{sec:conc}. 

\section{Background}
\label{sec:back}

\subsection{Models}

The excitation energy and angular momentum of a fission fragment  are dissipated through neutron and photon emission. Neutron and photon emission in a given fission event is thus a correlated process. The event-by-event correlations are also expected to affect the neutron and photon multiplicities relative to each other within a given fission event~\cite{Nifenecker1972}.  Physics-based fission event generators have been developed to simulate the complex process of fission fragment excitation and de-excitation, We consider here two such fission event generators:  \texttt{CGMF}~\cite{Talou2013,Talou2016} and \texttt{FREYA}~\cite{Vogt2009,Vogt2011, Vogt2013, Vogt2014a, Vogt2014b}. A comprehensive discussion of the fission models used in \texttt{CGMF} and \texttt{FREYA} is given in Ref.~\cite{Talou2018} and the references therein. In both fission event generators, the neutron-photon multiplicity correlation is not an input parameter but rather a prediction arising from the simulated excitation mechanisms. The fission event generators simulate energy and spin (and parity in the case of \texttt{CGMF}) of the fission fragments as the fragments de-excite to their ground states by emission of neutrons and photons.

\begin{figure}[h!]
\centering
\includegraphics[width= 0.4\textwidth]{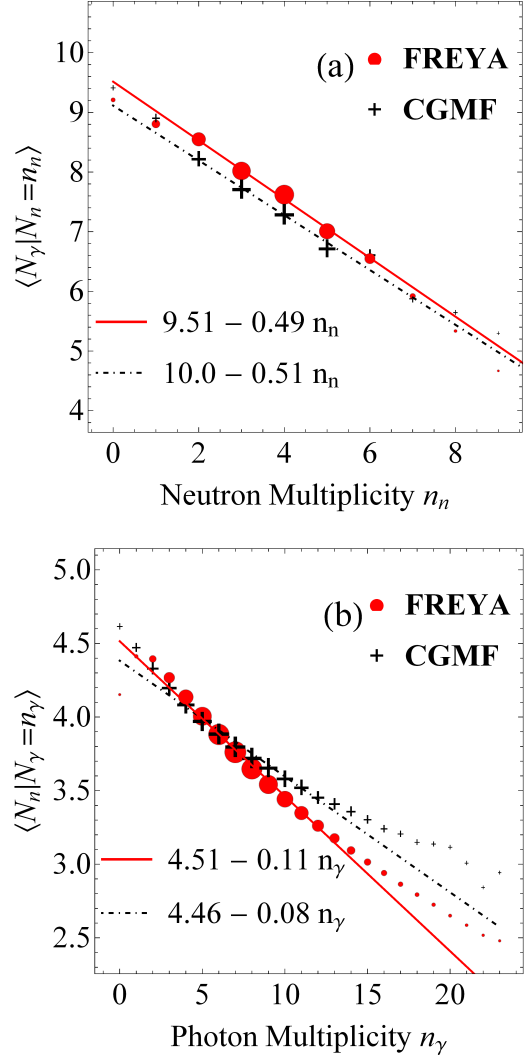}
\caption[Regression Plot]{Regression plots of simulated neutron and photon multiplicities from $^{252}$Cf for \texttt{CGMF} and \texttt{FREYA}, for both (a) mean photon multiplicity as a function of neutron multiplicity and (b) mean neutron multiplicity as a function photon multiplicity. Linear regression lines and their relations are shown in the legends of the plots. We note that the $y$ axes on both (a) and (b) are zero-suppressed and that the axis scales are chosen to better reflect the relatively smaller neutron multiplicity.}
\label{fig: regressions}
\end{figure}

Both \texttt{CGMF} and \texttt{FREYA} predict decreasing mean neutron (photon) multiplicity as the photon (neutron) multiplicity increases. The results of regression analysis of the simulated fission events from both generators are shown in Fig.~\ref{fig: regressions}. In Fig.~\ref{fig: regressions}(a), the simulated points are positioned at the conditional mean neutron multiplicity for each photon multiplicity while the size of each point is proportional to the relative probability of that photon multiplicity. The roles of neutron and photon multiplicities are reversed in Fig.~\ref{fig: regressions}(b). The least-squares linear fits indicated in the legends are shown in solid lines. A lower energy threshold of $150$ keV, chosen to match the energy thresholds of the experiments we analyze in the next section, is assumed here and for the rest of this work in the discussion of \texttt{CGMF} and \texttt{FREYA}. From the figure, it is apparent that both \texttt{CGMF} and \texttt{FREYA} predict negative neutron-photon emission correlations. The competition over the finite sources of energy and angular momentum of the fragment plays a determinant role in establishing these correlations. 

Both \texttt{CGMF} and \texttt{FREYA} predict a predominantly linear correlation. It is appropriate to describe their correlations by an event-by-event multiplicity covariance, $\text{cov}(N_n, N_\gamma)$:
\begin{subequations}
\label{eq: modelCov}
\begin{eqnarray}
\text{cov}_{\text{\texttt{FREYA}}} (N_n,N_\gamma) &= -0.8200 \pm 0.0004 \ , \\
 \text{cov}_{\text{\texttt{CGMF}}} (N_n,N_\gamma) &= -0.839 \pm 0.004 \ ,
\end{eqnarray}
\end{subequations}
where the uncertainties are purely due to limited run-time of the event generators. Uncertainties arising from the the models assumed in these generators are not included.  A higher amount of \texttt{FREYA} events were available, leading to a lower statistical uncertainty.

Although linearity in the regressions shown in Fig.~\ref{fig: regressions} is lost for events with high multiplicities, these events are very rare and have a negligible effect on the overall regression slope. The study of non-linear multiplicity correlations will be pursued in future work. 

\subsection{Experimental Results}

We note that several types of multiplicity correlations exist. In addition to the event-by-event analysis discussed here, neutron and photon emission has also been correlated through fragment-by-fragment analyses. Fragment-by-fragment analyses assign a mean neutron and a mean photon multiplicity to specific fission fragments, as determined over an ensemble of measured fission events. The mean multiplicities are then correlated with each other through the fragments \cite{Nifenecker1972}.  Thus, fragment-by-fragment analyses integrate over all possible de-excitation paths that a fragment may take. Event-by-event analyses, on the other hand, assign a neutron multiplicity and a photon multiplicity to each fission event, thereby retaining information on the de-excitation path followed by the fragment. Although both fragment-by-fragment and event-by-event analyses can produce correlations between neutron and photon multiplicities, they are two distinct types of correlations that cannot generally be expected to agree. This point is made clear in  Ref.~\cite{Glassel1989}: the same data produce a positive multiplicity correlation when analyzed fragment-by-fragment and a negative multiplicity correlation when analyzed event-by-event.

There have been previous experimental analyses of multiplicity correlations in the spontaneous fission of $^{252}$Cf. Experimental determinations of neutron-photon multiplicity correlations have been published by Nifenecker \textit{et al.}~\cite{Nifenecker1972}, Gl\"{a}ssel \textit{et al.}~\cite{Glassel1989, SchmidFabian1988}, Bleuel \textit{et al.}~\cite{Bleuel2010}, Wang \textit{et al.}~\cite{Wang2016}, and Marcath \textit{et al.}~\cite{Marcath2018}. The results reported in these papers, while open to interpretation, have been used to suggest the existence of a positive, negative, or effectively null multiplicity correlation between neutrons and photons. The studies by Nifenecker \textit{et al.} \cite{Nifenecker1972} and Wang \textit{et al.}~\cite{Wang2016} were performed on a fragment-by-fragment basis and are not relevant for this investigation and are thus not discussed further. The results of Gl\"{a}ssel \textit{et al.} \cite{Glassel1989, SchmidFabian1988}, Bleuel \textit{et al.} \cite{Bleuel2010}, and Marcath \textit{et al.} \cite{Marcath2018} were reported on an event-by-event basis. Gl\"{a}ssel \textit{et al.} determined a small negative correlation, while both Marcath \textit{et al.} and Bleuel \textit{et al.} reported a much smaller or null correlation. 

The results reported by these authors came in the form of regression slopes.  Marcath \textit{et al.} and Gl\"{a}ssel \textit{et al.} determined the regression slope of the mean emitted neutron multiplicity as a function of the photon multiplicity, $\alpha_n$. Conversely, Bleuel \textit{et al.} determined the regression slope of the mean emitted photon multiplicity as a function of the neutron multiplicity, $\alpha_\gamma$. We can convert the regression slopes to a covariance by
\begin{equation}
\text{cov}(N_n, N_\gamma) = \alpha_n \sigma^2(N_\gamma) = \alpha_\gamma \sigma^2(N_n) \ ,
\label{eq:recToCov}
\end{equation}
where $\sigma^2(N_n)$ and $\sigma^2(N_\gamma)$ are the variances of the neutron and photon multiplicity distributions. We use the neutron multiplicity  variance, $1.59 \pm 0.02$, as determined by Santi \textit{et al.} \cite{Santi2008a} and the photon multiplicity variance, $13.4 \pm 3.2$, as determined Oberstedt  \textit{et al.}\cite{Oberstedt2015}. \texttt{FREYA} and \texttt{CGMF} predict variances of $1.69$  and $1.81$ for neutrons, and $7.73$ and $10.7$ for photons. Although the results presented in this work are affected by the choice of variance, the qualitative conclusions are not. 

After converting the regression slopes to multiplicity covariances, we directly compare the experimental determinations with one another. The results are shown in Table~\ref{tab:discBefore}.  Presently, the experimental data does not lead to a unified conclusion on the existence of event-by-event neutron-photon multiplicity correlations. 
\begin{table}
\centering
\caption{\label{tab:discBefore}  Comparison of the covariances determined by the experimental determinations of an event-by-event multiplicity correlation found in the literature. More information can be found in the references provided for each experiment. }
\begin{tabular}{| l | l |}  	
\hline
 Experiment & $\text{cov}(N_n, N_\gamma)$ \\
\hline
Gl\"{a}ssel \textit{et al.}  \cite{Glassel1989} & $-0.35 \pm 0.01$  \\
\hline
Bleuel \textit{et al.} \cite{Bleuel2010} &  $0.0 $  \\
\hline
Marcath \textit{et al.} \cite{Marcath2018} &  $-0.02 \pm 0.12$  \\
\hline
\end{tabular}
\end{table}

\section{Unfolding Techniques}
\label{sec:unf}

The primary purpose of unfolding techniques used in this work is to determine the neutron-photon multiplicity covariance at emission, $\text{cov}(N_n, N_\gamma)$, from the multiplicity covariance between detected neutrons and photons, $\text{cov}(D_n, D_\gamma)$. 

We have developed new unfolding techniques for the analysis of event-by-event multiplicity data. Each of these techniques addresses an experimental bias that may affect the measurement. These biases are: background, particle misclassification, neutron-photon pileup, and inelastic photon production. The correction for these biases are performed on the neutron-photon multiplicity covariance directly and not on the probability distribution. Because the covariance can be computed from the linear regression of incomplete data, the corrections we have developed do not require a complete probability distribution. 

\subsection{Binomial Response Unfolding}

The analytic model of system response used in this work is based on the binomial-response model developed by Diven \textit{et al.}~\cite{Diven1956a} and further investigated by Nifenecker \textit{et al.}~\cite{nifeneckergroups}. See also the work by Bzdak \textit{et al.}~\cite{Bzdak2016, Bzdak2017} for more recent applications of the binomial-response model. The system response relates the emitted neutron and photon multiplicities, $N_n$ and $N_\gamma$, to the detected multiplicities, $D_n$ and $D_\gamma$. Binomial response assumes that every particle is detected independently of all other particles and that each particle is detected with a constant detection efficiency, $\epsilon_{n}$ and $\epsilon_\gamma$ for neutrons and photons, respectively. Neutron and photon detection efficiencies are expressed in terms of the emitted and detected mean multiplicities as \cite{Bzdak2016}
\begin{subequations}
\label{eq:epsDer}
\begin{eqnarray}
\langle D_n \rangle &=& \epsilon_n \langle N_n \rangle \ , \\
\langle D_\gamma \rangle &=& \epsilon_\gamma \langle N_\gamma \rangle \ .
\end{eqnarray}
\end{subequations}
In the binomial response model, covariance unfolding is performed by correcting for efficiency losses, \textit{i.e.},
\begin{equation}
\text{cov}_{\text{Bin}}(N_n, N_\gamma) = \frac{\text{cov}(D_n, D_\gamma)}{\epsilon_n \epsilon_\gamma} \ ,
\label{eq:baseBin}
\end{equation}
where the detection efficiencies, $\epsilon_n$ and $\epsilon_\gamma$, are computed using Eq.~\eqref{eq:epsDer} by comparing the detected mean multiplicities with published nuclear data for the mean neutron and photon multiplicities from the spontaneous fission of $^{252}$Cf \cite{Santi2008a, Oberstedt2015}. The unfolding given in Eq.~\eqref{eq:baseBin} does not, however, model several biases that can significantly alter the experimentally-determined covariance.

\subsection{Background}

Apart from external sources that can be reduced by an appropriate choice of shielding, the californium source is itself a significant source of positively correlated photons and neutrons from $\alpha$ decay and $(\alpha,n)$ reactions, respectively. It is important to experimentally determine the effects of background and remove this bias.

The background neutron and photon multiplicities, $B_n$ and $B_\gamma$ respectively, are modeled as completely uncorrelated with the spontaneously-fissioning source. Because background particles are indistinguishable from particles emitted following fission, the measured multiplicity is the sum of the two multiplicities, $D_n + B_n$ and $D_\gamma + B_\gamma$ respectively. Using the bilinearity property of covariance \cite{degrootProbability}, we find that the detected covariance should be replaced by
\begin{equation}
\text{cov}(D_n, D_\gamma) \rightarrow \text{cov}(D_n, D_\gamma) + \text{cov}(B_n, B_\gamma) \ .
\label{eq:back}
\end{equation}

\subsection{Misclassification}

The material properties of some detectors allow them to  be sensitive to both neutrons and photons and, in some cases, to distinguish the two particles. However, the assignment of a specific particle type to an interaction is not a perfect process, and some misclassification is expected. Particle misclassification provides a source of particles correlated to the fission source.

Let $\theta_n$ be the probability of a neutron being misclassified as a photon, and similarly for $\theta_\gamma$. The covariance in the detected multiplicities is then biased as
\begin{eqnarray}
\text{cov}(D_n,D_\gamma) &\rightarrow & \left[(1-\theta_n)(1 - \theta_\gamma) + \theta_n \theta_\gamma \right] \text{cov}(D_n,D_\gamma)\nonumber \\
&  \ & + \theta_n(1- \theta_n) \left[\sigma^2(D_n) - \langle D_n \rangle \right] \nonumber \\
&  \ & + \theta_\gamma(1- \theta_\gamma) \left[\sigma^2(D_\gamma) - \langle D_\gamma \rangle \right] \ ,
\label{eq:misc}
\end{eqnarray}
where $\sigma^2(D_n)$ and $\sigma^2(D_\gamma)$ are the variances of the detected neutron and photon distributions. The misclassification bias simplifies significantly in the regime $\epsilon_{n,\gamma} \ll 1$,  for either particle. The relationship
\begin{equation}
\sigma^2(D_{n,\gamma}) = \langle D_{n, \gamma} \rangle + \mathcal{O}(\epsilon_{n, \gamma}^2) 
\end{equation}
is applicable when the efficiency is sufficiently low. If we further assume that misclassification is small and identical for neutrons and photons,
\begin{equation}
\theta_n = \theta_\gamma = \theta \ll 1,
\end{equation}
the misclassification bias tends to
\begin{equation}
\text{cov}(D_n,D_\gamma) \rightarrow  (1-2 \theta) \ \text{cov}(D_n,D_\gamma) \ .
\label{eq:miscSimple}
\end{equation}

\subsection{Neutron-Photon Pileup}

Because the number of detectors in any detection system is finite and detectors cannot resolve and analyze interactions occurring in quick succession, the detection of particles effectively depletes the total number of available detectors for subsequent detections. This behavior, which cannot be modeled by the binomial response model, is called pileup. We note that the pileup bias persists whether the pileup event is removed by eliminating the entire event or by labeling it as a single interaction. 

Pileup biases the multiplicity covariance since the detection of a photon directly affects and decreases the probability of a neutron detection. This effect is dependent on the number of detectors $k$ and is calculated to be \cite{MarinINMM2019}
\begin{equation}
\text{cov}(D_n, D_\gamma) \rightarrow \frac{k}{k-1} \text{cov}(D_n,D_\gamma) - \epsilon_n \epsilon_\gamma \frac{ \langle N_n \rangle \langle N_\gamma \rangle }{k - 1} + \mathcal{O}(\epsilon_n^2, \epsilon_\gamma^2)  \ .
\label{eq:pileup}
\end{equation}
\subsection{Inelastic Photon Production}

The production of secondary particles by emitted neutrons and photons can be a significant source of positive bias in the determined covariance. This effect is dependent on the detector material and on the structure surrounding the source. Given the energies of photons emitted following fission, ($\gamma$, $n$) reactions are negligible. Inelastic photon production in neutron scattering, ($n$, $n' \gamma$) and ($n$, $\gamma$), can be non-negligible. 

We model the inelastic photon bias as
\begin{equation}
\text{cov}(D_n,D_\gamma) \rightarrow \text{cov}(D_n,D_\gamma) + \xi_n \epsilon_n \sigma^2(N_n) \ ,
\label{eq:inel}
\end{equation}
where $\xi_n$ is the probability that an emitted neutron interacts with material surrounding the source or within the detector and produces a photon that is subsequently detected. We note that Eq.~\eqref{eq:inel} is valid only in the case where detection of neutrons and photon inelastic production are independent processes. The extension to a detection dependent $\xi_n$ is trivial, but not required in this work. 

\subsection{Energy biasing}

We note here that the only explicit variables in the binomial-response model are the emitted multiplicities, the detected multiplicities, and the detection efficiencies. Therefore, the model misses some important underlying physical mechanisms. For example, the dependence of multiplicity on emission energy is not explicitly treated.  Here, we correct for the energy-dependent bias using a forward model of unfolding, \textit{i.e.}, showing that the experimentally-determined neutron-photon correlation is consistent with an assumed initial model. We perform this type of analysis in Section~\ref{sec:exp} when we analyze the Marcath \textit{et al.} data \cite{Marcath2018}.

\section{Experimental Results}
\label{sec:exp}

\subsection{Gl\"{a}ssel et al.}

Gl\"{a}ssel \textit{et al.}~\cite{Glassel1989} used the Darmstadt-Heidelberg Crystal Ball consisting of 162 sodium-iodide detectors in a $4\pi$ configuration~\cite{HeildelbergSphere}. An ion chamber was used to measure fission fragments emitted from a $^{252}$Cf(sf) source. The signal from the ion chamber was used as a fission time tag. An event-by-event determination of the mass and excitation energy of the fission fragments was also possible. Photons following fission were measured in the sodium iodide detectors, predominantly through full energy deposition. Inelastic scattering of neutrons on the sodium-iodide detectors, and the subsequent detection of the the inelastic photons, was the main mechanism of neutron detection. The distinction of neutron and photon events was made purely on the basis of time of flight. The effects of delayed prompt photons, which would appear as photons to the detectors, is quantified and removed \cite{SchmidFabian1988}.

\begin{figure}[h!]
\centering
\includegraphics[width= 0.4\textwidth]{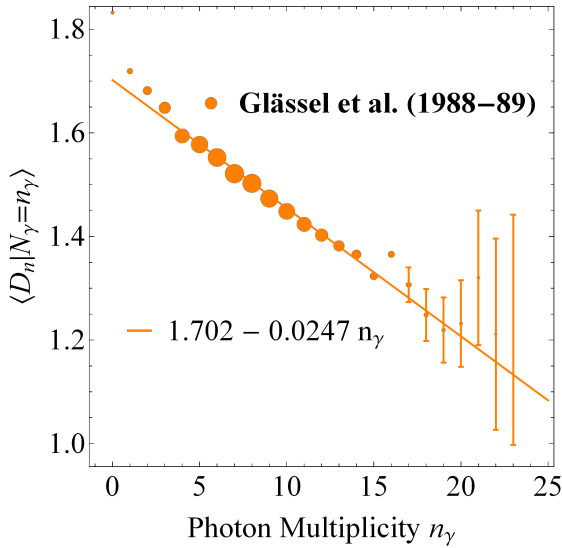}
\caption[graphGlassel]{Weighted linear regression analysis performed on the Gl\"{a}ssel \textit{et al.} data \cite{Glassel1989, SchmidFabian1988}. The marker size indicates the probability of the particular photon multiplicity, from Valentine and Oberstedt \textit{et al.}~\cite{Valentine2001, Oberstedt2015}. Note again that the $y$~axis is zero-suppressed.}
\label{fig: glasFig}
\end{figure}

Gl\"{a}ssel \textit{et al.} determined a small negative event-by-event correlation in the multiplicity of neutrons and photons from $^{252}$Cf. We have examined the original data published in Ref.~\cite{SchmidFabian1988} with cooperation from Gl\"{a}ssel and Schmid-Fabian, and present our analysis in Fig.~\ref{fig: glasFig}. Using the linear regression shown in Fig.~\ref{fig: glasFig} and the assumed emitted photon distribution from Ref.~\cite{Oberstedt2015}, we find a mean detected neutron multiplicity of $\langle D_n \rangle_{\text{Gl\"{a}ssel}} = 1.50 \pm 0.01$. Of course, this is much lower than the known experimental value of $\langle N_n \rangle = 3.75  \pm 0.01$ (\textit{e.g.}, from Ref.~\cite{Santi2008a}), indicating that the result was not corrected for neutron efficiency. The neutron efficiency we estimate using Eq.~\eqref{eq:epsDer}, $\approx 40 $ \%, is lower than the nominal neutron efficiency of the Crystal Ball, $\approx 60$ \%. However, pileup and the rejection of detections from neighboring detectors can reduce the neutron efficiency. The photon efficiency, nominally $>98 $ \%, does not need to be unfolded since it would only represent a negligible contribution.

In the analysis of the Gl\"{a}ssel \textit{et al.} data presented in Ref. ~\cite{SchmidFabian1988}, Schmid-Fabian estimates pileup to account for $15$ \% of the correlation in the measured data. Our own perturbative pileup correction, Eq.~\eqref{eq:pileup} with higher-order terms considered, predicts a similar estimate of the impact of photon-neutron pileup, approximately $ 14$ \%. 

The corrected covariance between detected multiplicities is then finally corrected for neutron efficiency by using the binomial response, Eq.~\eqref{eq:baseBin} with neutron and photon efficiencies of $0.4$ and $0.98$ respectively. We find
\begin{equation}
\text{cov}_{\text{Gl\"{a}ssel}}(N_n,N_\gamma) \approx -0.71 \pm 0.17 \ .
\label{glassel}
\end{equation} 
%


\subsection{Bleuel et al.}

The result reported in Bleuel \textit{et al.} \cite{Bleuel2010} suggests that the multiplicity correlation in $^{252}$Cf is small or null for at least some portion of the fission fragment yields. Bleuel \textit{et al.} measured event-by-event correlations using the LiBerACE~\cite{liberace2010} detector at Lawrence Berkeley National Laboratory. The system comprised 6 HPGe clover detectors in face-centered cubic configuration, each clover surrounded by 16 BGO detectors. The system has a high efficiency for gamma detection,  $>50$ \%, and $4\pi$ solid angle coverage. Furthermore, using the excellent energy resolution of the HPGe detectors and the background suppression techniques presented in Ref.~\cite{Bleuel2010}, the isotopic identity of certain fragments could be determined on an event-by-event basis. 

The mean photon multiplicity regression on the neutron multiplicity was computed by measuring the photon multiplicity distribution from fragment pairs with different neutron multiplicities, \textit{i.e.}, $N_n = 2$ for  $^{106} \text{Mo} + ^{144}\text{Ba}$, and $N_n = 4$ for $^{106} \text{Mo} + ^{142}\text{Ba}$. Bleuel \textit{et al.} did not measure neutrons directly, effectively eliminating the issue of photon-neutron pileup and misclassification. 

The LiBerACE system employed by Bleuel \textit{et al.} is insensitive to direct neutron interactions. However, the combination of the germanium high neutron-inelastic scattering cross section and a high photon efficiency lead to a large inelastic photon bias. In collaboration with Bleuel \textit{et al.}, we have been able to identify the presence of inlelastic photon production in the collected data. In an MCNPX-\texttt{PoliMi} \cite{Pozzi2003, Pozzi2012a} simulation of the array we have determined a value of $0.35$ for the inelastic photon production probability, $\xi_n$, with neutron energies sampled from the fission spectrum \cite{Watt1952a, Pozzi2003}. Because Bleuel \textit{et al.} could identify the post-evaporation fragments, the neutron efficiency  $\epsilon_n$ is unity. Using Eq.~\eqref{eq:inel} to correct for this bias, and correcting for the resultant expression for photon efficiency using Eq.~\eqref{eq:baseBin}, we find
\begin{equation}
\text{cov}_{\text{Bleuel}}(N_n, N_\gamma) \approx -0.69 \pm 0.10 \ .
\end{equation}
%


We also note that the energy threshold used by Bleuel \textit{et al.}, 100 keV, is slightly lower than the 150 keV threshold used in CGMF and FREYA. However, we have found through FREYA calculations that the dependence of the multiplicity correlation on the photon energy threshold is small in this region. In fact, FREYA predicts a slightly more negative covariance, $\approx -0.90$, at $100$ keV than it does at $150$ keV.

\subsection{Marcath et al. }

We analyze the data collected and reported by Marcath \textit{et al.}~\cite{Marcath2018}. They measured the neutron-photon emission of $^{252}$Cf using the Chi-Nu array at Los Alamos National Laboratory.  The Chi-Nu array comprises up to 54 EJ-309 liquid organic scintillators, each $17.78$ cm in diameter and $5.08$ cm thick. These detectors are distributed on a hemispherical surface centered on the fissioning source at a mean distance of $1.05 \pm 0.02$ m. During the measurement, 42 detectors were active. The overall solid angle coverage provided by the 42 detectors was $0.083 \times 4\pi$ sr, with intrinsic efficiencies of $\epsilon_n \approx 32$\% and $\epsilon_\gamma  \approx 23$\%. A fission chamber was used to signal the occurrence of a spontaneous fission event within the sample. Because of its high sensitivity to virtually all the properties of the emitted particles, including energy, timing, angular distribution, and multiplicity, the system suffered from substantial systematic bias. 

Background is measured for a fixed time interval before the fission chamber trigger. The covariance of the background is estimated to represent approximately $15$ \% of the determined covariance. The misclassification rate, $\theta$, is determined through time of flight and pulse shape discrimination. A conservative value of $0.01$ is used for $\theta$ \cite{Marcath2018}. The effects of pileup are dominant in Chi-Nu. With 42 detectors, we estimate the pileup bias to represent approximately $67 \% $ of the determined covariance. Lastly, due to time cuts applied to particle detection, the effects of inelastically-produced photons were reduced experimentally. 

After removal of these biases and corrections for efficiencies, the covariance was determined to be
\begin{equation}
\text{cov}_{\text{Marcath}}(N_n,N_\gamma) = -0.58 \pm 0.06 \ .
\end{equation} 

\begin{figure}[h!]
\centering
  \includegraphics[width= 0.4\textwidth]{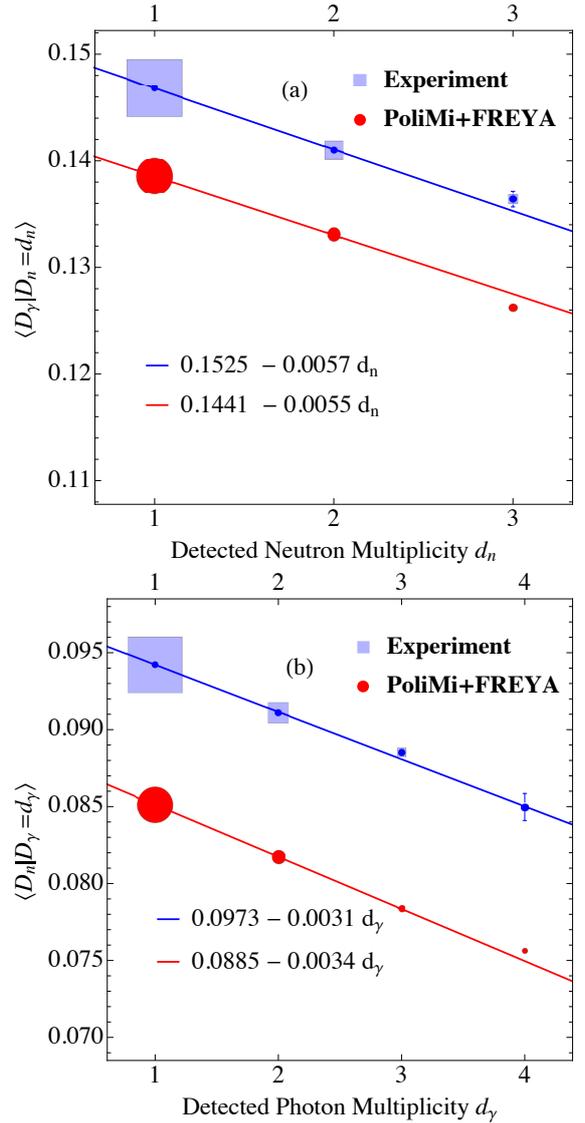}
\caption{Regression plots of neutron and photon multiplicity from $^{252}$Cf for our experiment, for both (a) mean photon multiplicity as a function of neutron multiplicity and (b) mean neutron multiplicity as a function photon multiplicity. We note that the $y$ axes on both (a) and (b) are zero-suppressed and that the scales on the $y$ axes are different. The results are compared with the MCNPX-\texttt{PoliMi} simulation of the Chi-Nu array using \texttt{FREYA} as the simulated source of neutrons and photons. }
\label{fig:marcFig}
\end{figure}

As mentioned in Section~\ref{sec:unf}, the efficiency correction of Eq.~\eqref{eq:baseBin} does not inherently model the energy dependence of the detector response. The effect of energy biasing in Marcath \textit{et al.} is significant, due to the variation of the efficiency of the organic scintillators at different incident energies. To check the effects of this bias, we perform an MCNPX-\texttt{PoliMi} simulation, using \texttt{FREYA} events to simulate the particles from the fission source. This forward unfolding model allows us to determine whether \texttt{FREYA} is consistent with the experimental observation. The result of the \texttt{PoliMi}+\texttt{FREYA} simulations are compared to the experimental results in Fig.~\ref{fig:marcFig}. The covariance that would have been determined experimentally if \texttt{FREYA} were the correct model of emission is $-0.57\pm 0.01$, in good agreement with the corrected covariance we determined for the Marcath \textit{et al.} data. Therefore, we conclude that the results of Marcath \textit{et al.} are consistent with the \texttt{FREYA} calculations. The discrepancy in the intercepts between the linear regressions of \texttt{FREYA} and Marcath \textit{et al.} data, which is on the order $5-6$ \%, can be attributed to an underestimation of the system efficiency in simulation.

Furthermore, we have analyzed the bias due to energy dependent efficiencies, as well as that due to partial energy deposition by photons on the organic scintillators. After inverting the photon system response we have qualitatively confirmed, independently of fission event generators, that the results of Marcath \textit{et al.} are consistent with a negative covariance of $\approx -0.90 \pm 0.15 $. However, this result is affected by large statistical uncertainties. 

\section{Conclusion}
\label{sec:conc}

\subsection{Discussion}

The results of our analysis are summarized in Table~\ref{tab:discFinal}. In the same table, we have also show the covariances predicted by \texttt{CGMF} and \texttt{FREYA}. When comparing Tables~\ref{tab:discBefore} and \ref{tab:discFinal}, we see that the inconsistency between experimental results, and between experiment and models, is qualitatively resolved. 

\begin{table}[htbp]
\caption{\label{tab:discFinal} Summary of the discussion of Section~\ref{sec:disc}. For each of the  reported values, both model based calculations and experimental determinations, we provide our determination of the corrected covariance.
}
\centering
\begin{threeparttable}
\begin{tabular}{|l | l | }  	
\hline
 Determination & $\text{cov}(N_n, N_\gamma)$  \\
\hline
\texttt{CGMF} ~\cite{Talou2013,Talou2016} & $-0.839 \pm 0.004$  \\
\hline
\texttt{FREYA}~\cite{Vogt2009,Vogt2011, Vogt2013, Vogt2014a, Vogt2014b} & $-0.8200 \pm 0.0004$  \\
\hline
Gl\"{a}ssel \textit{et al.}~\cite{Glassel1989, SchmidFabian1988} & $-0.71 \pm 0.17$ \\
\hline
Bleuel \textit{et al.}~\cite{Bleuel2010} & $-0.69 \pm 0.10 $  \\
\hline
Marcath \textit{et al.}~\cite{Marcath2018} & $-0.58 \pm 0.06 $ \tnote{a} \\
\hline
\end{tabular}
        \begin{tablenotes}
            \item[a] Energy sensitivity considerations show that this result is also in quantitative agreement with \texttt{FREYA}. 
        \end{tablenotes}
 \end{threeparttable}
\end{table}

Although Gl\"{a}ssel \textit{et al.} and Bleuel \textit{et al.} indicate a slightly smaller correlation than \texttt{CGMF} and \texttt{FREYA}, the unfolding of these results was limited and some secondary effects were neglected. In particular, background was not addressed in our analysis of either of these experiments. Furthermore, the uncertainties associated with \texttt{CGMF} and \texttt{FREYA} in Table~\ref{tab:discFinal} are purely statistical, and are likely underestimated in this work. Both the input parameters and the nuclear de-excitation models employed in these event generators carry uncertainties not quantified in this work. It is then entirely possible that these experimental results, after more careful error analysis, will also be in good quantitative agreement with the fission event generators.

\subsection{Summary and Future Work}
We have presented evidence for the existence of a negative event-by-event correlation in neutron and photon multiplicities in the prompt emission following the spontaneous fission of $^{252}$Cf. Although the published experimental results on neutron-photon event-by-event correlations were previously not in agreement with one another or with model calculations, we show in this work that, after  experimental considerations and corrections, they are all in agreement. This agreement provides an experimental validation of the models used by the fission event generators \texttt{CGMF} and \texttt{FREYA}. In future experiments, it will be important to detect fragments in coincidence with neutrons and photons. By isolating mass and excitation energy of fragments, it will be possible to study in detail the competition of neutrons and photons during the de-excitation process. The relationship between intrinsic excitation energy and angular momenta, and its effects on neutron-photon multiplicity correlations, will be the subject of future work.

Aside from the results presented here, the question of a possible correlation between momenta and multiplicities, for both neutrons and photons, remains. Some work has already been performed \cite{Lecouteur1959, Lemaire2006, Schuster2019}, but no definitive conclusion has been reached. Although determining these correlations would represent an important milestone in the characterization of fission signatures from $^{252}$Cf(sf), the same treatment should be applied to other isotopes undergoing spontaneous fission, notably $^{240}$Pu. A more tailored unfolding analysis should also be applied to isotopes undergoing neutron-induced fission, especially at higher incident energies where multi-chance fission and pre-equilibrium contributions would most likely complicate these correlations. The complexity introduced by incident neutrons is a necessary obstacle to be overcome, given the importance of induced-fission reactions for applications involving the multiplication properties of fission \cite{Mueller2016, Shin2017, Enqvist2010}.

\section{Acknowledgements}

S. M. would like to thank Prof. P. Gl\"{a}ssel and Dr. R. Shmid-Fabian for providing a copy of the dissertation published by the latter in 1987. S. M. is thankful to Prof. P. Gl\"{a}ssel  and Dr. D. Bleuel for discussions of their experimental results. S.M. would also like to acknowledge the contributions from Isabel E. Hernandez, Eoin P. Sansevero, and Dr. Tony H. Shin to the analysis of the experimental data, and Dr. J\o rgen Randrup for discussion of the physical implications of our results in FREYA. This work was in part supported by the Office of Defense Nuclear Nonproliferation Research \& Development (DNN R\&D), National Nuclear Security Administration, US Department of Energy. This research was funded in-part by the Consortium for Verification Technology under Department of Energy National Nuclear Security Administration award DE-NA0002534. The work of V.A.P.  was performed under the auspices of UT-Battelle, LLC under Contract No. DE-AC05-00OR22725 with the U.S. Department of Energy. The work of R.V. was performed under the auspices of the U.S. Department of Energy by Lawrence Livermore National Laboratory under Contract DE-AC52-07NA27344. The work of P.T. and M.J.M. was performed under the auspices of the National Nuclear Security Administration of the U.S. Department of Energy at Los Alamos National Laboratory under Contract DE-AC52-06NA25396. 

\bibliography{mybib}

\end{document}